\documentclass[prl,twocolumn,superscriptaddress,showpacs,aps]{revtex4-1}

\usepackage{color}
\usepackage{graphicx}
\usepackage{bm}
\usepackage{amsmath}

\begin{document}

\newcommand{\Ce}{CeCoIn$_5$}
\newcommand{\Tc}{T_{\text c}}
\newcommand{\Hcii}{H_{\text c2}}
\newcommand{\Hq}{H_Q}
\newcommand{\Ho}{H_{\text 0}}
\newcommand{\dwave}{d_{x^2-y^2}}
\newcommand{\Tm}{TmNi$_2$B$_2$C}
\newcommand{\Er}{ErNi$_2$B$_2$C}
\newcommand{\Lu}{LuNi$_2$B$_2$C}
\newcommand{\Y}{YNi$_2$B$_2$C}

\title{Vortex lattice studies in {\Ce} with $H \perp c$}

\author{P. Das}
\affiliation{Department of Physics, University of Notre Dame, Notre Dame, IN 46556, USA}

\author{J. S. White}
\affiliation{School of Physics and Astronomy, University of Birmingham, B15 2TT, United Kingdom}
\affiliation{Laboratory for Neutron Scattering, Paul Scherrer Insitut, CH-5232 Villigen, Switzerland}

\author{A. T. Holmes}
\affiliation{School of Physics and Astronomy, University of Birmingham, B15 2TT, United Kingdom}

\author{S. Gerber}
\affiliation{Laboratory for Neutron Scattering, Paul Scherrer Insitut, CH-5232 Villigen, Switzerland}

\author{E. M. Forgan}
\affiliation{School of Physics and Astronomy, University of Birmingham, B15 2TT, United Kingdom}

\author{A. D. Bianchi}
\affiliation{D\'epartement de Physique and RQMP, Universit\'e de Montr\'eal, Montr\'eal, QC H3C 3J7, Canada}

\author{M. Kenzelmann}
\author{M. Zolliker}
\affiliation{Laboratory for Developments and Methods, Paul Scherrer Institut, CH-5232 Villigen, Switzerland}

\author{J. L. Gavilano}
\affiliation{Laboratory for Neutron Scattering, Paul Scherrer Insitut, CH-5232 Villigen, Switzerland}

\author{E. D. Bauer}
\author{J. L. Sarrao}
\affiliation{Los Alamos National Laboratory, Los Alamos, NM 87545, USA}

\author{C. Petrovic}
\affiliation{Brookhaven National Laboratory, Upton, NY 11973, USA }

\author{M. R. Eskildsen}
\email{eskildsen@nd.edu}
\affiliation{Department of Physics, University of Notre Dame, Notre Dame, IN 46556, USA}

\date{\today}

\begin{abstract}
We present small angle neutron scattering studies of the vortex lattice (VL) in {\Ce} with magnetic fields applied parallel ($H$) to the antinodal $[100]$ and nodal $[110]$ directions. For $H \parallel [100]$, a single VL orientation is observed, while a $90^\circ$ reorientation transition is found for $H \parallel [110]$. For both field orientations and VL configurations we find a distorted hexagonal VL with an anisotropy, $\Gamma = 2.0 \pm 0.05$. The VL form factor shows strong Pauli paramagnetic effects similar to what have previously been reported for $H \parallel [001]$. At high fields, above which the upper critical field ($\Hcii$) becomes a first-order transition, an increased disordering of the VL is observed.
\end{abstract}


\maketitle

The heavy fermion $d$-wave superconductor {\Ce} continues to attract great interest due to the fascinating interplay between unconventional superconductivity and magnetism exhibited by this material~\cite{Petrovic01,Izawa01,Tayama02,Bianchi02,Bianchi03b}. In particular {\Ce} satisfies the stringent conditions believed to be necessary for the existence of a spatially modulated superconducting order parameter first described by Fulde, Ferrell, Larkin and Ovchinnikov (FFLO)~\cite{FFLOa,FFLOb}: a first order superconducting transition at low temperatures indicating Pauli limiting~\cite{Bianchi03a} and a mean free path much greater than the superconducting coherence length~\cite{Kasahara05}. While a number of bulk measurements support a FFLO phase in {\Ce}~\cite{Radovan0304,Bianchi03a,Matsuda07}, an experimental verification by a direct probe of the predicted spatial variation of superconductivity has not yet been reported. It is important because its unusual order parameter has relevance to a wide range of different areas of physics ranging from magnetars/neutron stars~\cite{Casalbuoni04}, ultracold atomic gases~\cite{Zwierlein06} to organic superconductors~\cite{Lortz07}. Furthermore, a field induced incommensurate antiferromagnetic order ($Q$-phase) is observed in the same high field - low temperature region of the phase diagram where the FFLO state is proposed to exist~\cite{Kenzelmann08,Kenzelmann10,Blackburn10}. At present the detailed connection between the FFLO and the $Q$-phase remains an area of active research~\cite{Hiragi10,Curro10,Koutroulakis10,Kumagai11,Suzuki11}.

Information about the nature of the superconducting state in {\Ce} can be obtained from studies of the vortex lattice (VL). Previous VL studies with $H \parallel [001]$ using small angle neutron scattering (SANS) have shown a rich vortex phase diagram and strong Pauli paramagnetic effects~\cite{Eskildsen03,Schmitt06,Bianchi08,Kawamura08,White10}, which extend well below the field $\Ho$ where $\Hcii$ becomes a second order transition~\cite{Bianchi03a}. In this field orientation, the supercurrents and quasiparticles travelling in the plane perpendicular to $H$ are strongly affected by the $\dwave$ nodes of the superconducting order parameter and a square VL aligned with the nodes is formed over a significant portion of the VL phase diagram~\cite{Bianchi08,White10}. For $H \perp c$, the symmetry in the plane $\perp H$ is no longer tetragonal and nodal effects are less strong, though Pauli-limiting effects are similar in magnitude for all field directions.

In this letter we report the results of SANS studies of the VL in {\Ce} with the magnetic field applied in the basal plane which are the first such studies in any $d$-wave superconductor. Experiments were performed for field applied both along the antinodal ($[100]$) and nodal ($[110]$) directions~\cite{Izawa01,An10,Weickert06,Vorontsov10,Hiragi10}. For both field orientations a distorted hexagonal VL is observed, reflecting the anisotropy in the $ac$-plane. A single VL orientation is observed at all fields for $H \parallel [100]$, while the VL for $H \parallel [110]$ undergoes a $90^\circ$ reorientation transition in the range $7.5 - 8.7$~T. Measurement of the VL form factor show a strong Pauli paramagnetic effect similar to what we have previously reported for $H \parallel [001]$. A broadening of the VL rocking curves is observed at fields above $\Ho$, suggesting that the abrupt vortex nucleation associated with a first-order $\Hcii$ leads to a freezing in of disordering. Further experimental details are provided elsewhere~\cite{White10,SOM}.

\begin{figure}
  \includegraphics{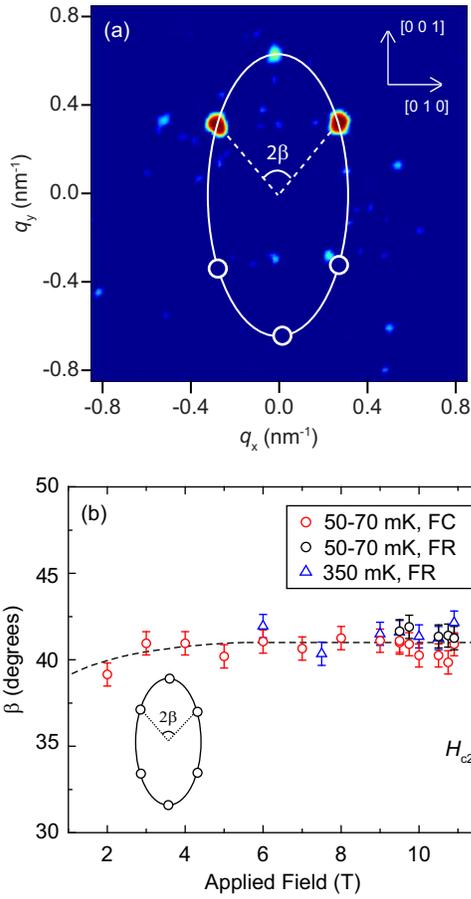}
  \caption{(Color online)
           VL structure with $H \parallel [100]$ (anti-nodal direction). (a) Diffraction pattern obtained at 9.0~T and 50~mK showing a distorted hexagonal VL. Only Bragg peaks on the upper half of the detector were rocked through the Ewald sphere. The open circles show positions of Bragg peaks obtained by a reflection through $q = 0$. (b) Field dependence of the VL opening angle $\beta$ defined in (a), at 50 and 350~mK, following either a field cool (FC) or field ramp (FR) procedure~\cite{SOM}. The dotted line represents $\Hcii$ at 50~mK~\cite{Bianchi03a}.
           \label{Fig1}}
\end{figure}

We first discuss the symmetry and orientation of the VL, beginning with $H \parallel [100]$. Fig.~\ref{Fig1}(a) shows a VL diffraction pattern with the peak positions fitted to an ellipse, indicating a distorted hexagonal VL.
The distortion can be quantified by the VL opening angle $\beta$. As shown in Fig.~\ref{Fig1}(b) the VL structure remains unchanged for all measured fields and temperatures, with a constant opening angle, $\beta = 41.0^\circ \pm 0.7^\circ$ at high fields, and perhaps a slight fall at low fields. The opening angle is directly related to axial ratio of the ellipse in Fig.~\ref{Fig1}(a), which is a measure of the anisotropy in the plane perpendicular to the applied field, $\Gamma_{100} = \tan 60^\circ / \tan \beta = 1.99 \pm 0.05$. This quantity mainly reflects the anisotropy of the Fermi velocity in the $ac$-plane but includes gap effects, especially at higher fields~\cite{Hiragi10}.

\begin{figure}[t!]
  \includegraphics{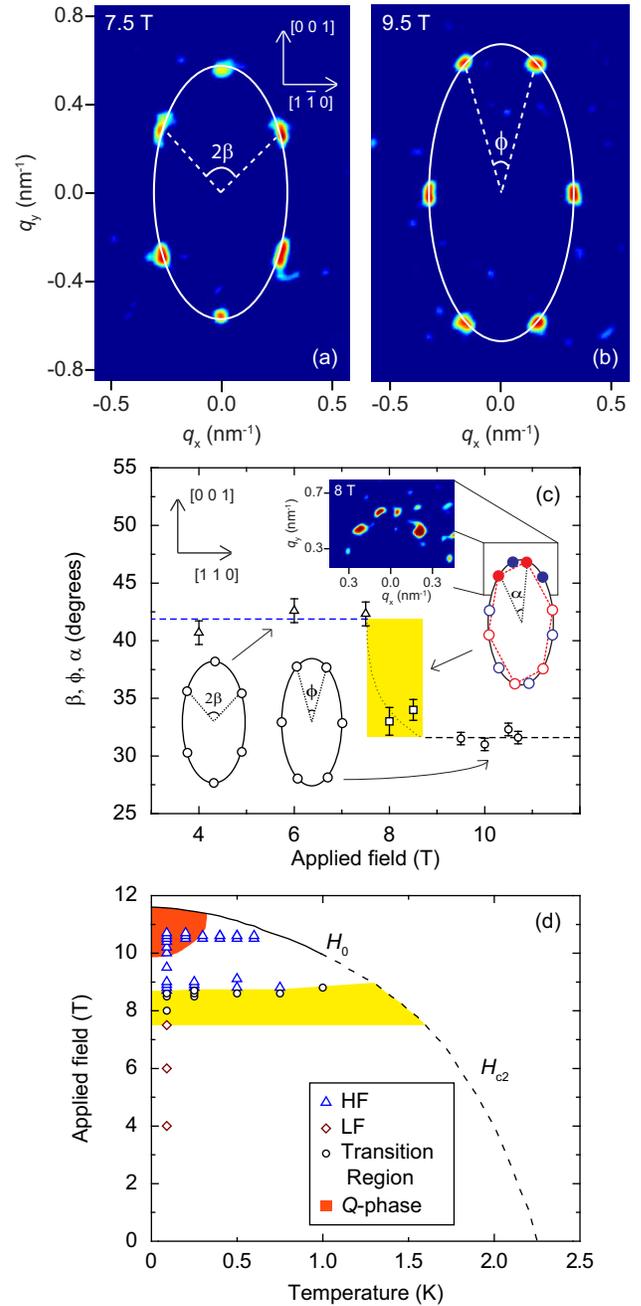}
  \caption{(Color online)
           VL with $H \parallel [110]$ (nodal direction). Diffraction pattern obtained at 90~mK and at an applied field of $7.5$~T (a) and $9.5$~T (b) respectively. (c) Field dependence of the VL opening angles $\beta$, $\phi$ and $\alpha$ at 90~mK. The yellow area indicates the transition region between the two VL orientations. The inset shows a diffraction pattern as well as a schematic of the rotated two-domain HF-phase. (d) VL phase diagram showing the extent of the LF and HF VL phases. $\Ho$ and $\Hcii$ are adapted from ref.~\onlinecite{Bianchi03a} and $Q$-phase from ref.~\onlinecite{Kenzelmann08}.
           \label{Fig2}}
\end{figure}

Rotating the sample such that $H \parallel [110]$ significantly changes the VL phase diagram. At low field a distorted hexagonal VL is observed, similar to that found for $H \parallel [100]$, with Bragg peaks on the major axis of the ellipse as shown in Fig.~\ref{Fig2}(a). However as the field is increased the VL undergoes a $90^\circ$ reorientation transition to a configuration with Bragg peaks on the minor ellipse axis as shown in Fig.~\ref{Fig2}(b). We denote these two VL orientations as respectively the low-field (LF) and high-field (HF) phases, with the former observed for $H \leq 7.5$~T and the latter above $8.7$~T. In between these two fields a VL rotation away from the HF-phase is observed as shown in the inset to Fig.~\ref{Fig2}(c), but is difficult to study systematically due to weak scattering in the transition region. Fig.~\ref{Fig2}(c) shows the field dependence of the VL opening angles. The opening angle $\alpha$, defined in the inset, remains close to $\phi$, suggesting that the transition to the LF-phase is first-order. Within the LF and HF phases the opening angle again remains essentially field-independent with $\beta = 41.9^\circ \pm 1.0^\circ$ and $\phi = 31.6^\circ \pm 0.5^\circ$, corresponding to $\Gamma^{\text {LF}}_{110} = 1.94 \pm 0.07$ and $\Gamma^{\text {HF}}_{110} = \tan 30^\circ / \tan (\phi/2) = 2.05 \pm 0.03$. While $\Gamma_{100}, \Gamma^{\text {LF}}_{110}$ ~and~ $\Gamma^{\text {HF}}_{110}$ could be considered identical within the experimental error, we note that the relative values are in qualitative agreement with the theoretical prediction of Hiragi {\em et al.}~\cite{Hiragi10}:
$\Gamma_{100} = \Gamma^{\text {LF}}_{110} < \Gamma^{\text {HF}}_{110}$
(although this was for a field of $0.211 \Hcii \approx 2.4$~T).

The VL phase diagram with $H \parallel [110]$ is summarized in Fig.~\ref{Fig2}(d). Similar to what was found for $H \parallel [001]$~\cite{Bianchi08,White10}, the transition does not coincide with the appearance of the magnetic $Q$-phase~\cite{Kenzelmann08,Kenzelmann10, Blackburn10} nor the proposed FFLO-phase~\cite{FFLOa,FFLOb,Radovan0304,Bianchi03a,Matsuda07}. More importantly, our results demonstrate a distinct difference between the antinodal and nodal directions in \Ce. In contrast to the VL anisotropy, $\Gamma$, our phase diagram does not agree with the theoretical predictions of opposite VL orientation transitions with increasing temperature for fields along $[100]$ and $[110]$ respectively~\cite{Hiragi10}. This is not surprising since calculations of the VL structure depend sensitively on the Fermi surface and gap anisotropy and are notoriously difficult due to the small energy differences involved. This illustrates the need to include realistic material parameters in such calculations.

It is interesting to compare the VL behavior to that reported for \Lu \ and \Er \ with $H \parallel [100]$, where a $90^\circ$ VL first order reorientation is observed~\cite{Eskildsen01a}. Thermal conductivity measurements in non-magnetic \Lu \ and \Y \ have suggested the presence of point-nodes along the $\langle 100 \rangle$ directions~\cite{Maki02,Izawa02}, indicating that the VL reorientation is observed in cases where the field is applied along the nodal direction and is unrelated to a vortex core expansion which is believed responsible for the reentrance of the square VL phase in {\Ce} with $H \parallel [001]$~\cite{White10}.

\begin{figure}
  \includegraphics{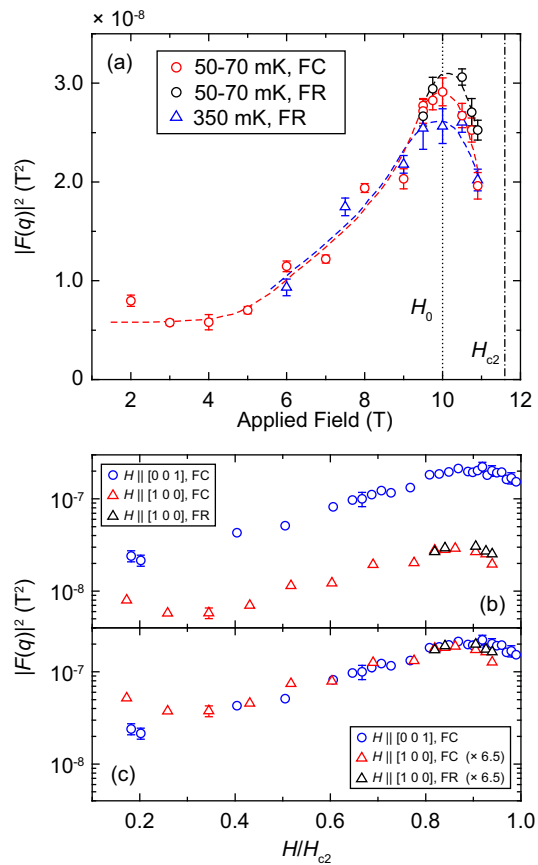}
  \caption{(Color online)
           Field dependence of VL form factor. (a) Form factor with $H \parallel [100]$. The dashed lines are guides to the eye and $\Hcii$ is shown for
           $T = 50$~mK. (b) Comparison of $|F(q)|^2$ measured with field along $[100]$ and $[001]$~\cite{White10}. (c) Multiplying the form factor for $H \parallel [100]$ by $(2.55)^2 = 6.5$ causes $|F(q)|^2$ for the two field orientations to fall on a single curve.
           \label{Fig3}}
\end{figure}

\begin{figure}
  \includegraphics{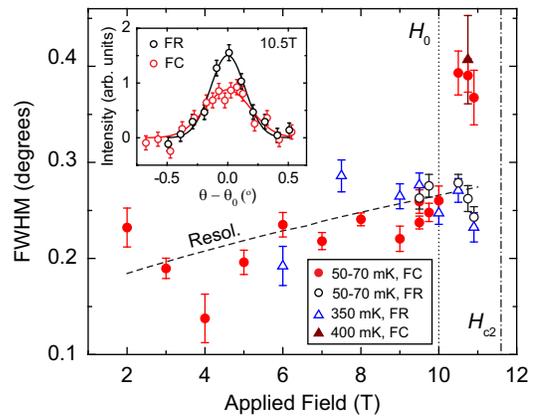}
  \caption{(Color online)
           VL rocking curve widths corresponding to the data in Fig 1(b). The significant scatter in the data at low
           fields is due to the low scattered intensity and higher background at shorter $q$. The calculated experimental resolution is shown by the dashed line. The insert compares the rocking curves obtained at 10.5~T following a FC and FR. The lines are Gaussian fits to the data and $\theta_0$ denotes the center of the rocking curve.
           \label{Fig4}}
\end{figure}

We now turn to studies of the VL form factor $F(q)$, which is a measure of the magnetic field modulation in the mixed state. To avoid complications arising from a changing VL orientation, and the difficulty obtaining a well-ordered single domain VL in the transition region, these measurements were performed only for $H \parallel [100]$. The form factor is related to the integrated reflectivity $R$ of the VL, which is obtained by rotating and/or tilting the cryomagnet and sample such that the VL scattering vectors cut through the Ewald sphere. Examples of rocking curves obtained in this fashion are shown in the inset to Fig.~\ref{Fig4}. Gaussian fits to the rocking curves give an integrated intensity which is divided by the incident neutron flux to yield the integrated reflectivity
\begin{equation}
  R=\frac{2\pi \gamma^2 \lambda_{\text n}^2 t}{16 \phi_0^2 q } \left| F(q) \right|^2,
\end{equation}
where $\gamma = 1.913$ is the magnetic moment of the neutron in nuclear magnetons, $\lambda_{\text n}$ is the neutron wavelength, $t$ is the sample thickness, $\phi_0 = h / 2e = 2067$~T~nm$^2$ is the flux quantum and $q$ is the length of the VL scattering vector. The resulting VL form factor for $H \parallel [100]$ is shown in Fig.~\ref{Fig3}(a).
Contrary to the commonly observed exponential decrease with increasing field~\cite{Densmore09}, $|F(q)|^2$ increases with increasing field up to a maximum at 10~T after which it decreases on approaching the upper critical field. We note that a qualitatively similar behavior is observed at 50 and 350~mK and for different VL preparation (FC/FR). However there are subtle differences between the form factor above $\Ho = 10$~T when one compares VLs obtained following a FC and a FR. We will return to this point later.

The unusual field dependence of the form factor is similar to what has been reported for {\Tm}~\cite{Schmitt07} and {\Ce} with $H \parallel [001]$~\cite{White10,Bianchi08,Schmitt06}, and can at least partly be explained by Pauli paramagnetic effects~\cite{Schmitt07,Ichioka07,Mineev10}. Briefly, there is a significant spin-polarization of the unpaired quasi-particles in the vortex cores, resulting in an increased amplitude of the modulation of the magnetic field. Moreover, it was recently shown that accounting for antiferromagnetic fluctuation in the vortex cores yields both a skewing and an increase of the form factor~\cite{Ikeda10,Aoyama11}, in better agreement with our findings. One notes that while the maximum in $|F(q)|^2$ coincides with the onset of the $Q$-phase at 50~mK, the two effects do not appear to be linked. This is supported by the weak $T$-dependence of the field at which the form factor is maximum and that a similar behavior is observed for $H \parallel [001]$~\cite{White10} where no $Q$-phase has been observed. Moreover, the temperature dependence of the VL peak intensity (top of rocking curve) at $10.9$~T showed no features except an abrupt drop to zero at $\Hcii$.

Fig.~\ref{Fig3}(b) and (c) shows a comparison of the form factors obtained with the applied field perpendicular and parallel to the crystalline basal plane as a function of the reduced field~\cite{Hciinorm}, which can be collapsed onto a single curve by multiplying $|F_{100}|$ by a constant factor of $2.55 \pm 0.1$. As we have previously shown~\cite{White10}, the form factor at $\Hcii$ is directly related to the jump in the magnetization, yielding $\Delta M_{001}/\Delta M_{100} \approx 2.8$~\cite{Tayama02} which is in quite good agreement with the scaling of $|F_{100}|$. What is remarkable is that the rescaled form factors agree as a function of reduced field \textit{over the entire measured range}, whereas one would expect the ratio to gradually approach the Fermi velocity anisotropy ($\sim 2$) at low fields where Pauli paramagnetic effects are small. This scaling invites detailed theoretical analysis.

Finally we return to the observation that above $\Ho$ the VL acquires a field/temperature history dependence, leading to a difference between the FC or a FR case. As shown in Fig.~\ref{Fig3}(a), the measured form factor is lower following a FC than following a FR for fields above 10~T, indicating a more well-formed VL in the latter case. The same trend is seen in Fig.~\ref{Fig4} showing the field dependence of the rocking curve widths, which is inversely proportional to the longitudinal correlation length of the VL. Below 10~T the width is resolution limited. Above $\Ho$, one sees a striking increase in the FC rocking curve widths. We speculate that this is due to the abrupt appearance of the vortices while slowly cooling through the first-order superconducting transition, immediately freezing in of any disordering. Furthermore, the VL can be effectively annealed by inducing vortex motion (FR), as shown in the inset in Fig.~\ref{Fig4} which compares FC and FR rocking curves for 10.5~T.

In summary we have studied the VL in {\Ce} with $H \perp c$. While a single, distorted hexagonal VL is found for $H \parallel [100]$, a $90^\circ$ reorientation is observed for $H \parallel [110]$. The VL form factor shows Pauli paramagnetic effects, similar to what is observed for $H \parallel [001]$. Finally both the form factor and the rocking curve widths show a clear dependence on the VL preparation above $\Ho$.

We acknowledge support from the US NSF through grant DMR-0804887, the EPSRC of the UK, the Alfred P Sloan Foundation, NSERC (Canada), FQRNT (Qu\'ebec), the Canada Research Chair Foundation, the Swiss National Centre of Competence in Research program 'Materials with Novel Electronic Properties', and from the European Commission under the 6th Framework Programme through the Key Action: Strengthening the European Research Area, Research Infrastructures, Contract No. RII3-CT-2003-505925. Work at Los Alamos was performed under the auspices of the US DOE. Part of this work was carried out at the Brookhaven National Laboratory, which is operated for the US Department of Energy by Brookhaven Science Associates (DE-Ac02-98CH10886).



\begin{thebibliography}{99}
\bibitem{Hciinorm}
  The field is divided by the measured upper critical field for $H \parallel [100]$ and $H \parallel [001]$. Alternatively one could normalize to the extrapolated orbital critical field~\cite{Tayama02}. Since $H_{{\text c2},100}^{\text {orb}}/H_{{\text c2},001}^{\text {orb}} =2.33 \approx H_{{\text c2},100}^{\text {meas}}/H_{{\text c2},001}^{\text {meas}} = 2.34$ this will merely change the value on the $x$-axis in Figs.~\ref{Fig3}(b) and (c) but not the comparison.
\end{thebibliography}
\end{document}